\documentstyle[aps,prb,twocolumn,floats,psfig]{revtex}

\begin{document}

\bibliographystyle{prsty}

\title{
Structure of a superconducting vortex pinned by a screw
dislocation \vspace{-1mm} }

\author{
Eugene M. Chudnovsky  }

\address{
Department of Physics and Astronomy, Lehman College, City University of New York, \\
250 Bedford Park Boulevard West, Bronx, New York 10468-1589 \\
\smallskip
{\rm(Received 26 June 2001)}
\bigskip\\
\parbox{14.2cm}
{\rm Spatial dependence of the magnetic field and the
superconducting current in a flux line pinned by a screw
dislocation are computed. Interaction of a superconducting vortex
with the chiral-symmetry breaking elastic strain of a screw
dislocation results in a helical current along the axis of the
dislocation. It is argued that screw dislocations make impossible
a force-free arrangement of flux lines in the presence of a
transport current.
\smallskip
\begin{flushleft}
\end{flushleft}
} }
\maketitle

Pinning of vortices by screw dislocations can be of interest
because of the experimental evidence of this effect in Y-Ba-Cu-O.
\cite{Gerber,Hawely,Diaz,Dam}
Studies of pinning by dislocations are almost as old as the
Abrikosov's idea of superconducting vortices (see, e.g., the
review by Campbell and Evetts \cite{CampEvet} and references
therein).
More recently, these studies have been revived in
connection with high-temperature superconductors. \cite{GurPas}
They are normally based upon the expansion of scalar parameters of
a superconductor, such as, e.g., $T_{c}$, in terms of the
deformation tensor and its trace.

A superconducting vortex coupled to a screw dislocation is an
interesting theoretical problem, because of the broken chiral
symmetry of the deformation field produced by such a dislocation.
Same as the magnetic field ${\bf B}$, a screw dislocation running,
e.g., along the Z-axis, is an axial vector, say ${\bf p}$, that is
invariant under reflection.
Consequently, $p={\bf p}{\cdot}{\hat{\bf z}}={\pm}1$ is a
pseudoscalar that changes sign under reflection.
This makes possible a linear relation between the polar vector of
the current density ${\bf j}$ and the axial vector of the magnetic
field ${\bf B}$: ${\bf j}\,{\propto}\,p\,{\bf B}$, the relation
that would be otherwise prohibited by the invariance with respect
to reflections.

A closely related problem of a vortex coupled to a spiral defect
has been studied by Ivlev and Thompson \cite{IvThom1,IvThom2} for
an extreme case of a layered superconductor with a Josephson
coupling between the layers.
In such case, a spiral defect, running perpendicular to the
layers, geometrically connects them by a continuous helical path
around the defect.
Ivlev and Thompson elegantly solved this problem in spiral
coordinates and, in accordance with the above symmetry arguments,
demonstrated the existence of a ``fountain-like" current along the
axis of the defect.

The purpose of this paper is to solve the problem in the opposite
extreme case of a flux line coupled to a screw dislocation in an
isotropic three-dimensional superconductor.
In such case the existence of longitudinal currents parallel to
the dislocation core is somewhat less obvious.
The model we suggest is complementary to the model of Ivlev and
Thompson.
It is based upon the description of dislocations within continuous
elastic theory.
We find that superconducting currents do flow along screw
dislocations in three-dimensional superconductors, although the
spatial dependence of these currents is different from the one
found in Ref.7.
Our results should be directly relevant to MgB$_2$ and
conventional isotropic superconductors.
The Y-Ba-Cu-O, where the coupling between vortices and screw
dislocations has been experimentally observed, falls somewhere
between the two models.

The free energy of the system is ${\cal{F}}={\cal{F}}_{D}+
{\cal{F}}_{N}+{\cal{F}}_{GL}$, where ${\cal{F}}_{D}$ is the energy
of the dislocation, ${\cal{F}}_{N}$ is the energy of the normal
electron liquid in the absence of the magnetic field, and
${\cal{F}}_{GL}$ is the Ginzburg-Landau free energy,
\begin{eqnarray}\label{GL}
{\cal{F}}_{GL} & = & {\int}d^{3}r\left[\frac{({\bf
\nabla}{\times}{\bf A})^ {2}}{{8\pi}}\right] + \nonumber
\\
& & {\int}d^{3}r\left[a|\psi|^{2}+\frac{b}{2}|\psi|^{4} +
\frac{{\hbar}^{2}}{4}(D_{i}{\psi})m^{-1}_{ij}(D_{j}{\psi})^{*}\right].
\end{eqnarray}
Here ${\bf A}$ is the vector potential (${\bf B}={\bf
\nabla}{\times}{\bf A}$),  ${\psi}=|{\psi}|\exp(i{\phi})$ is the
complex order parameter of the superconducting phase, $a$ and $b$
are constants, $D_{i}$ is the gauge-invariant derivative,
\begin{equation}\label{longD}
D_{i}={\nabla}_{i}-\frac{2ie}{{\hbar}c}A_{i},
\end{equation}
and $m_{ij}$ is the tensor of effective masses.
For an isotropic superconductor, in the absence of crystal
defects, $m_{ik}=m{\delta}_{ik}$.

The presence of a dislocation results in a non-zero elastic
strain,
\begin{equation}\label{deftensor}
u_{ij} = \frac{1}{2}({\nabla}_{i}u_{j}+{\nabla}_{j}u_{i}).
\end{equation}
At distances exceeding a few lattice spacings from the dislocation
core, the components of the dimensionless tensor $u_{ij}$ are
small \cite{LL} and the parameters of the superconductor, such as
$a$,$b$, and $m_{ij}$, can be expanded into the power series of
$u_{ij}$.
We shall see that for a screw dislocation $Tr(u_{ij})=0$, that is,
screw deformations change the symmetry of the crystal but not the
local density.
Thus, to the lowest order in $u_{ij}$, the interaction of the
screw dislocation with the Ginzburg-Landau order parameter can be
introduced by the substitution
\begin{equation}\label{transm}
m{\delta}_{ij} \;\;{\longrightarrow}\;\; m({\delta}_{ij}+gu_{ij}),
\end{equation}
where $g$ is a dimensionless parameter.
From the physics of the electron states in crystals, it is clear
that the effect of the lattice deformations on the tensor of
effective masses is small as long as $u_{ij}$ are small.
If some of $u_{ij}$ are large, the effect should be also large.
Consequently, $g$ must be of order unity.

Since crystal defects are insensitive to superconductivity,
${\cal{F}}_{D}$ must have a very weak dependence on ${\psi}$.
We will neglect that dependence.
Then the variation of ${\cal{F}}_{GL}$ with respect to
${\psi}$ and ${\bf A}$ gives the following two Ginzburg-Landau
equations:
\begin{equation}\label{GL1}
-\frac{{\hbar}^{2}}{4m}g_{ij}D_{i}D_{j}{\psi}+a{\psi}+b|\psi|^{2}{\psi}=0
\end{equation}
and the Maxwell equation ${\bf \nabla}{\times}{\bf B} =
\frac{4\pi}{c}{\bf j}$ with
\begin{equation}\label{GL2}
j_{i} =
-g_{ij}\left[\frac{ie{\hbar}}{2m}({\psi}^{*}{\nabla}_{j}{\psi}
-{\psi}{\nabla}_{j}{\psi}^{*})+
\frac{2e^{2}}{mc}A_{j}|\psi|^{2}\right],
\end{equation}
where $g_{ij}={\delta}_{ij}+gu_{ij}$.

Eq.\ (\ref{GL1}) describes the vortex core where the concentration
of Cooper pairs, $|\psi|^{2}$, changes from zero at the center of
the vortex to a constant value, $|a|/b$, at distances exceeding
the coherence length ${\xi}={\hbar}/2(m|a|)^{1/2}$.
At such distances the second Ginzburg-Landau equation becomes
\begin{equation}\label{London}
{\lambda}_{L}^{2}g^{-1}_{ij}({\bf \nabla}{\times}{\bf B})_{j}
+A_{i} = \frac{{\Phi}_{0}}{2{\pi}}{\nabla}_{i}\phi,
\end{equation}
where $\phi$ is the phase of $\psi$, $\;{\Phi}_{0}=hc/2e$ is the
flux quantum, and
${\lambda}_{L}=(mc^{2}/8{\pi}e^{2}|\psi|^{2})^{1/2}$ is the London
penetration depth.
We shall consider the case of ${\lambda}_{L}>>{\xi}$ which is
relevant to high-temperature superconductors.
At a large distance from the vortex core, where ${\bf B}$
exponentially goes to zero, Eq.\ (\ref{London}) reduces to ${\bf
A} = ({\Phi}_{0}/2\pi){\bf \nabla}{\phi}$.
After the integration over a closed distant contour enclosing the
vortex, it produces the conventional condition of the quantization
of the magnetic flux.
This condition remains unchanged by the deformations.
Applying curl to both sides of Eq.\ (\ref{London}), one obtains a
modified London equation
\begin{equation}\label{LonMod}
{\lambda}_{L}^{-2}B_{i}+{\epsilon}_{ijk}{\nabla}_{j}g^{-1}_{kl}({\bf
\nabla}{\times}{\bf B})_{l}=0.
\end{equation}

We can now turn to the problem of a flux line centered at a screw
dislocation parallel to the Z-axis of the crystal.
It should be naturally studied in circular cylindrical
coordinates: $z,r,{\varphi}$.
At distances, $r$, exceeding a few lattice spacings from the core
of the screw dislocation, the only non-zero component of the
displacement field ${\bf u}$ is \cite{LL}
\begin{equation}\label{u}
u_{z}=pb\frac{\varphi}{2\pi},
\end{equation}
where $p={\pm}1$ is the chirality of the screw dislocation and $b$
is the Burgers vector that coincides with the lattice spacing in
the Z-direction.
Consequently, the only non-zero components of the elastic strain
are
\begin{equation}\label{strain}
u_{z{\varphi}}=u_{{\varphi}z}=p\frac{b}{4{\pi}r}.
\end{equation}
Although the linear elastic theory fails near the axis of the
dislocation, for the purpose of our study the above formulas are
exact as long as the coherence length ${\xi}$ and the London
penetration depth ${\lambda}_{L}$ are greater than $b$.
As is shown below, the significant effect of the dislocation on
the superconducting vortex comes from the fact that the elastic
strain given by Eq.\ (\ref{strain}) is long range.

It is convenient to introduce a dimensionless coordinate
${\rho}=r/{\lambda}_{L}$ and a dimensionless small parameter
${\beta}=gb/4{\pi}{\lambda}_{L}$.
Substituting Eq.\ (\ref{strain}) into Eq.\ (\ref{LonMod}), and
retaining terms of the lowest order in ${\beta}$, one obtains the
following two equations for non-zero components of the magnetic
field $B_{z}(\rho)$ and $B_{\varphi}(\rho)$:
\begin{eqnarray}\label{twoeq}
B_{z} & = &
\frac{1}{\rho}\frac{d}{d\rho}\left[{\rho}\frac{dB_{z}}{d\rho}+\frac{p\beta}{\rho}\frac{d}{d\rho}({\rho}B_{\varphi})\right],
\\
B_{\varphi} & = &
\frac{d}{d\rho}\left[\frac{1}{\rho}\frac{d}{d\rho}({\rho}B_{\varphi})+
\frac{p\beta}{\rho}\frac{dB_{z}}{d\rho}\right].
\end{eqnarray}
For a flux line parallel to Z, the condition ${\beta}<<1$ results
in $B_{\varphi}<<B_{z}$ for all $r>{\xi}$.
Consequently, the effect of the dislocation on Eq.(11) can be
neglected and the system of equations for $B_{z}(\rho)$ and
$B_{\varphi}(\rho)$ can be solved by iteration.
Putting ${\beta}=0$ in Eq.(11) one obtains a conventional solution
for the Z-component of the magnetic field inside the flux line,
\begin{equation}\label{Bz}
B_{z}(\rho)=\frac{{\Phi}_{0}}{2{\pi}{\lambda}_{L}^{2}}K_{0}(\rho),
\end{equation}
where $K_{0}$ is the modified Bessel function and the coefficient
in front of $K_{0}$ is such that the total magnetic flux through
the XY-plane equals ${\Phi}_{0}$.
Then Eq.(12) can be reduced to the following form
\begin{equation}\label{BphiEq}
\frac{d}{d\rho}\left({\rho}\frac{dB_{\varphi}}{d\rho}\right)-\left({\rho}+\frac{1}{\rho}\right)B_{\varphi}=
-\frac{{p\beta}{\Phi}_{0}}{2{\pi}{\lambda}_{L}^{2}}K_{2}(\rho).
\end{equation}
The solution of this equation, that goes to zero at
$r{\rightarrow}\infty$, is
\begin{equation}\label{Bphi}
B_{\varphi}(\rho)=\frac{{p\beta}{\Phi}_{0}}{4{\pi}{\lambda}_{L}^{2}}\left[CK_{1}(\rho)+K_{1}(\rho)\ln(\rho)-
\frac{1}{\rho}K_{0}(\rho)\right],
\end{equation}
where $C$ is a constant of integration (to be computed later).

The current density in the flux line is given by ${\bf
j}=(c/4\pi){\bf \nabla}{\times}{\bf B}$.
The ${\varphi}$-component of ${\bf j}$ is the conventional vortex
current
\begin{equation}\label{Jphi}
j_{\varphi}(\rho)=-\frac{c}{4{\pi}{\lambda}_{L}}\frac{dB_{z}}{d{\rho}}=
\frac{c{\Phi}_{0}}{8{\pi}^{2}{\lambda}_{L}^{3}}K_{1}(\rho).
\end{equation}
The unusual feature of the problem, as in Ref.7, is the presence
of the Z-component of the current,
\begin{equation}\label{genJz}
j_{z}(\rho)  =
\frac{c}{4{\pi}{\lambda}_{L}}\frac{1}{\rho}\frac{d}{d\rho}({\rho}B_{\varphi}).
\end{equation}
Substituting here $B_{\varphi}$ of Eq.\ (\ref{Bphi}), we obtain
\begin{equation}\label{Jz}
j_{z}(\rho)=\frac{{p\beta}c{\Phi}_{0}}{16{\pi}^{2}{\lambda}_{L}^{3}}\left[\frac{2}{\rho}K_{1}(\rho)
-K_{0}(\rho)\ln(\rho)-CK_{0}(\rho)\right].
\end{equation}
Note that the dependence of $j_{z}$ on $\rho$, given by Eq.\
(\ref{Jz}), differs from the one obtained by Ivlev and Thompson.
\cite{IvThom1}

The magnetic field at distances
significantly exceeding ${\lambda}_{L}$ is not possible in a
superconductor.
Consequently, the total superconducting current $I_{z}$, that
flows parallel to the dislocation outside the vortex core, must be
zero,
\begin{equation}\label{Iz}
I_{z}=\int^{\infty}_{1/{\kappa}}j_{z}(\rho)\,2{\pi}{\rho}\,d{\rho}=0,
\end{equation}
where we have introduced ${\kappa}={\lambda}_{L}/{\xi}$.
With the help of Eq.\ (\ref{Bphi}) and Eq.\ (\ref{Jz}), this
condition allows one to obtain constant $C$:
\begin{equation}\label{C}
C=\ln(\kappa)+{\kappa}K_{0}(1/\kappa)K_{1}^{-1}(1/\kappa).
\end{equation}

\begin{figure}[t]
\unitlength1cm
\begin{picture}(11,6.2)
\centerline{\psfig{file=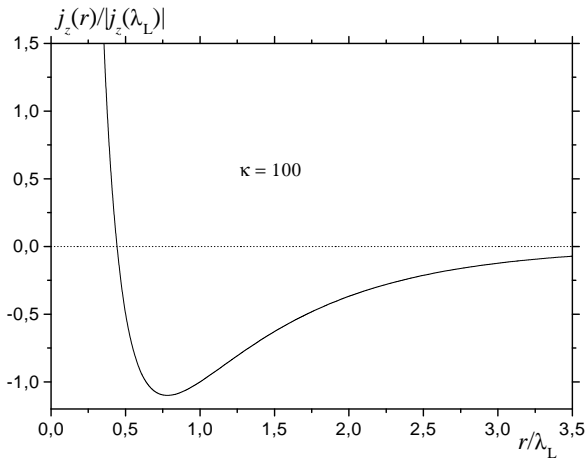,angle=-90,width=9cm}}
\end{picture}
\caption{ \label{fig_current} Radial dependence of the
longitudinal current density in a superconducting vortex coupled
to a screw dislocation.}
\end{figure}

A few observations are in order.
The $\kappa$ dependence of $j_{z}(\rho)$ is rather weak: $C$
changes from $C=0.1916$ at ${\kappa}=1/\sqrt{2}$ (which is the
boundary of type-II superconductivity) to the asymptotic form:
$C=2\ln(\kappa)$, at ${\kappa}\,{\rightarrow}\,{\infty}$.
For ${\kappa}=100$ and $p=1$, $\;j_{z}(r)$ is shown in Fig.\
\ref{fig_current}.
The maximal longitudinal current,
$j_{z}\,{\sim}\,(b/{\xi})j_{0}\;$ (with
$j_{0}=c{\Phi}_{0}/12{\sqrt{3}}{\pi}^{2}{\xi}{\lambda}_{L}^{2}$
being the Ginzburg-Landau critical current) occurs at
$r\,{\sim}\,{\xi}$ near the core of the vortex.
The longitudinal current $j_{z}$ changes sign at $r$ of order
${\lambda}_{L}$.
At $r>>{\lambda}_{L}$ it becomes exponentially small.
For a screw dislocation of the opposite chirality, the current is
in the opposite direction.

A screw dislocation either ends at the surface of the crystal or
forms a loop inside the crystal.
In the first case, the boundary condition,
$n_{i}g_{ij}D_{j}{\psi}=0$ (with $\;{\bf n}$ being the unit vector
normal to the boundary), prohibits currents through the surface of
the superconductor.
This results in a surface current that flows outward from the
vortex core at ${\xi}<r<{\lambda}_{L}$.
In the case of a dislocation loop, the flux line will follow the
loop.

The coupling of the flux line to a screw dislocation is the
$(b/\xi)^{2}$ fraction of the vortex-core energy available for
pinning.
In high-temperature superconductors, where the coherence length
can be of order of the Burgers parameter, dislocations can provide
a rather strong pinning \cite{GurPas}, in accordance with
observations \cite{Gerber,Hawely,Diaz,Dam}.
One should keep in mind, however, that the presence of the
$B_{\varphi}$ component of the field in the flux line pinned by a
screw dislocation makes impossible the ``force-free" situation in
which the field is parallel to the transport current.
In that sense columnar pins with no chirality have the advantage
over screw dislocations.

I thank Lev Bulaevskii and Tolya Kuklov for useful discussions.
This work has been supported by the U.S. Department of Energy
through Grant No. DE-FG02-93ER45487.



\end{document}